\documentclass[twocolumn]{article} % default is 10pt
\def \columnsep{0.2in}
\def \textwidth{7.45in}
\usepackage{graphicx}
\usepackage{hyperref} %/shareopt/tetex/share/texmf/tex/latex/hyperref/hyperref.sty
\usepackage{balance}

\newlength{\halfwidth}
\begin{document}

\title{\vspace{-0.75in} %Sample Template for SEN
Genetic~Improvement @ ICSE 2020} %Workshop}

\author{
        \begin{tabular}[t]{ccc}%
              
 %Modifications:
 %WBL 21 Jul 2020 Add affilations
 %WBL  6 Jul 2020 List email addresses for ciculation of drafts
 %WBL  4 Jul 2020 start with contributions for far: JP, BRBruce
 %look in author.tex r1.6 for SEN example of latex tabular format

\href{http://www.cs.ucl.ac.uk/staff/W.Langdon/}
{William B. Langdon} & 
\href{http://web.eecs.umich.edu/~weimerw/}
{Westley Weimer} % <weimerw@umich.edu>,
&
\href{http://www.cs.ucl.ac.uk/staff/J.Petke/}
{Justyna Petke}  % <j.petke@ucl.ac.uk> Tue, Jul 14, 2020 at 4:41 PM
\\
& University of Michigan, USA & University College, London, UK
\\[1ex]
\href{https://oakland.edu/secs/directory/fredericks}
{Erik Fredericks} % <fredericks@oakland.edu> Tue, Jul 7, 2020 at 3:50 PM
               %Fri, Jul 17, 2020 at 2:08 PM
&
\href{http://coinse.kaist.ac.kr/members/seongmin/}
{Seongmin Lee}   %Explainable-GI.txt <bohrok@kaist.ac.kr> Thu, Jul 9, 2020 at 10:59 AM
&
\href{https://www.lancaster.ac.uk/people-profiles/emily-winter}
{Emily Winter}   %Text on HF <e.winter@lancaster.ac.uk> Thu, Jul 9, 2020 at 4:14 PM
               % Tue, Jul 14, 2020 at 4:01 PM, Fri, Jul 17, 2020 at 10:18 AM
\\
Oakland University, USA &
KAIST, Korea &
Lancaster University, UK
\\[1ex]
\href{https://turintech.ai/about-us/#Mike}
{Michail Basios} %text in Industrial GI <mike@turintech.ai> Fri, Jul 10, 2020 at 6:34 PM
&
\href{http://web.cs.iastate.edu/~mcohen/}
{Myra B. Cohen}  %myra_gi.txt <mcohen@iastate.edu> Sat, Jul 11, 2020 at 9:15 PM
&
\href{http://www.cs.ucl.ac.uk/staff/a.blot/}
{Aymeric Blot}   %blot.txt on benchmarking <a.blot@cs.ucl.ac.uk> Mon, Jul 13, 2020 at 1:20 AM
\\
TurinTech, London, UK &
Iowa State University, USA &
University College, London, UK
\\[1ex]
\href{http://cs.adelaide.edu.au/~markus/}
{Markus Wagner}  %txt+ref <markus.wagner@adelaide.edu.au> Mon, Jul 20, 2020 at 3:42 PM
&
\href{https://web.cs.ucla.edu/~b.bruce/}
{Bobby R. Bruce} % <bbruce@ucdavis.edu>,
&
\href{https://cs.kaist.ac.kr/people/view?idx=516&kind=faculty&menu=160}
{Shin Yoo}       % <shin.yoo@kaist.ac.kr>
\\
Adelaide University, Australia &
UC, Davis, USA & %University of California
KAIST, Korea
\\[1ex]
\href{https://www-users.cs.york.ac.uk/simos/}
{Simos Gerasimou} %blog instead GI and SEAMS <simos.gerasimou@york.ac.uk> Mon, Jul 6, 2020 at 8:11 AM
&
\href{https://aist.fh-hagenberg.at/index.php/en/team-2/oliver-krauss-2}
{Oliver Krauss}   % <Oliver.Krauss@fh-hagenberg.at> Tue, Jul 28, 2020 at 8:48 AM
&
\href{http://www-personal.umich.edu/~yhhy/}
{Yu Huang}        % <yhhy@umich.edu> Tue, Jul 7, 2020 at 9:33 PM
\\
University of York, UK
&
AIST, Austria  %University of Applied Sciences Upper Austria &
&
University of Michigan, USA
\\[1ex]
\href{https://www.cs.iastate.edu/people/michael-gerten}
{Michael Gerten}  % <mcgerten@iastate.edu> Wed, Jul 8, 2020 at 5:51 PM
&
&
\\
Iowa State University, USA 
&
&
\\

\multicolumn{3}{c}{\includegraphics[width=3.5in]{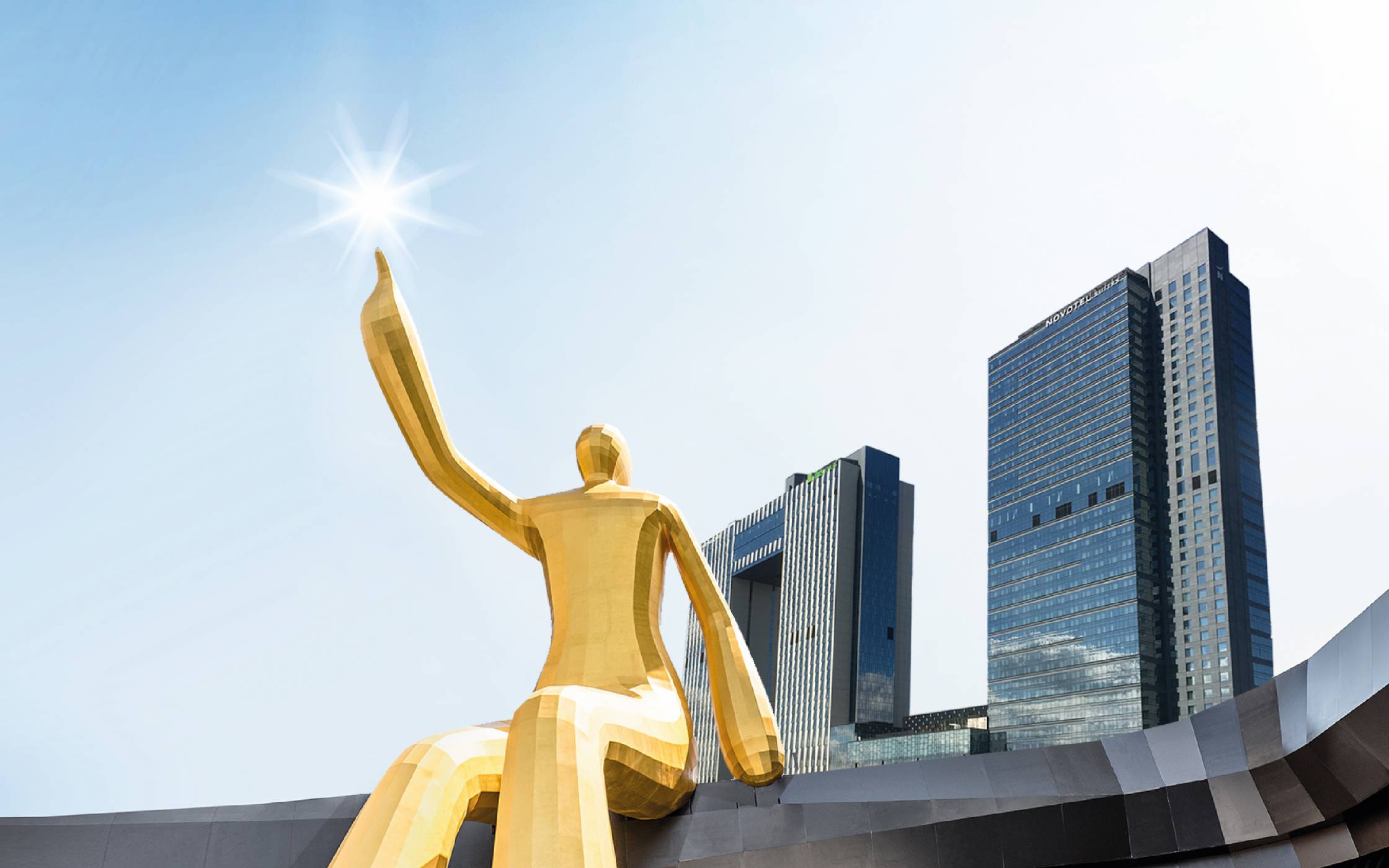}}\\
\\[2ex]
\multicolumn{3}{c}{Stand on the Shoulders of Giants
and Evolve} % Forward? from where we are?
        \end{tabular}%
}

\date{} %31 July 2020

\maketitle

\thispagestyle{empty} % turn off page numbers for SEN
\pagestyle{empty}     % turn off page numbers for SEN

\begin{abstract}
Following 
Prof.\ Mark Harman of Facebook's keynote 
and formal presentations
(which are recorded in the proceedings)
there was a wide ranging discussion at the eighth international
Genetic Improvement workshop, \mbox{GI-2020 @ ICSE}
(held as part of the
International Conference on Software
Engineering
on Friday 3$^{\rm rd}$~July 2020).
Topics included
industry take up,
human factors,
\mbox{explainabiloity}
(explainability, justifyability, exploitability)
and
GI benchmarks.
We also contrast various recent online approaches 
(e.g.\ SBST~2020) 
to holding 
virtual
computer science conferences and workshops 
via the WWW on the Internet
without face~to~face interaction.
Finally we speculate on
how the Coronavirus Covid-19 Pandemic will affect 
research next year and into the future.
\end{abstract}

\section{%Introduction: 
What is Genetic Improvement}

\href{https://en.wikipedia.org/wiki/Genetic_improvement_(computer_science)}
{Genetic Improvement}
is a branch
of Artificial Intelligence~(AI)
and Software Engineering
which applies optimisation %\cite{mhbj:manifesto}
to improve existing programs.
It is always possible to compare the new code
with the existing code
(effectively treating the program as its own specification)
allowing
GI to make measureable improvements to today's software.
Improvements may be functional
(e.g., does the new code have fewer bugs?
does it have a new feature?
does it give more accurate answers?)
or non-functional
(e.g.\ does it have better battery life?
is it more reliable?)

\vspace*{3ex} %fiddle p2 pagination
\section{GI @ ICSE 2020 via Zoom}

The eighth Genetic Improvement workshop (GI~2020) was held 
as part of the 
forty-second International Conference on Software Engineering
(ICSE~2020)
during the Corona virus Covid-19 pandemic.
Even as late as the close of ICSE workshop submissions, 
it was intended to hold ICSE in Seoul, the capital of South Korea,
from May~23 thru May~29, 2020.
However as the Pandemic bit, 
it became clear that May 2020 was not feasible.
Initially it was decided to delay the conference
(and hence the workshops) until October~2020.
This was later overturned as more experience with virtualising
conferences and holding them on the Internet was gained.
Hence, the 42$^{\rm nd}$~ICSE was held 
as a virtual electronic conference in early July 2020.
Some events were cancelled.
However
it was decided that
the 
GI~@~ICSE~2020 workshop would became an Internet only event
and was held using Zoom.us 
on Friday 3~July 2020 \mbox{13:00-16:20~UTC}
\url{http://geneticimprovementofsoftware.com/gi2020icse.html}
(Notice the Corvid imposed importance of the time zone.)

Fifty people pre-registered for the Genetic Improvement workshop.
On the day participation varied with about 35 people
``attending'' via Zoom
at any one time and
a further 40 or so watching on a live YouTube channel
(recording \url{https://youtu.be/GsNKCifm44A}).
Also 
Yu~Huang~@YuHuang\_yh
ensured that Twitter carried highlights on
\href{https://twitter.com/hashtag/gi_icse_2020?src=hashtag_click}
{\tt \#gi\_icse\_2020}.
As far as possible,
the workshop kept the traditional (physical) format,
starting with an invited keynote given by Prof.~Mark Harman,
who described the use of SBSE~\cite{mhbj:manifesto}
and Genetic Improvement 
\cite{White:2011:ieeeTEC}%
\cite{Alexander:2009:cec}% compiler code opt heuristic in GE
\cite{Langdon:2012:mendel}%
\cite{langdon:2015:hbgpa}%
\cite{Langdon:2013:ieeeTEC}%
\cite{Petke:2014:EuroGP}%Petke:2016:GPEM,%
\cite{Petke:2015:SBST}%
\cite{Petke:2017:ieeeTSE}%
\cite{Petke:gisurvey}%
\cite{Haraldsson:2020:GECCOcomp}% GI Tutorial,
,
within Facebook
and future plans, 
including social testing~\cite{Ahlgren:2020:GI}
and Facebook calls for research proposals.
This was followed by formal (albeit electronic only)
presentations of papers,
which are to be %ACM DL DOIs still broken 31 Jul 2020
published in the ACM digital library
\cite{%
Ahlgren:2020:GI,%
Blot:2020:GIsp,%
Blot:2020:GI,%
Dash:2020:GI,%
krauss_towards_2020,%
Winter:2020:GI}.
(The keynote and all of the 
presentations were recorded and are available via
YouTube
\url{https://youtu.be/GsNKCifm44A}).
The formal presentations were followed by free-form discussions
across seventeen time zones  
(again recorded).

In addition to Facebook,
industrial particpants
included 
\href{https://turintech.ai/}
{TurinTech}~(London UK)
and 
\href{https://news.grammatech.com}
{GrammaTech} (Bethesda MD and Ithaca NY, USA).
Academic and student particpants came from universities across
five continents.
(An edited record,
including genetic improvement tools and resources, 
of last year's workshop was
published in the 
ACM SIGSOFT's Software Engineering Notes \cite{Langdon:2019:SEN}
therefore we do not repeat that information here.)

Although the full discussion,
as recorded via Bobby Bruce's screen,
is available on YouTube,
the following sections condense more than an hour's
flowing discussion into just a few of the topics covered.

\begin{figure}[h]
\centerline{\includegraphics{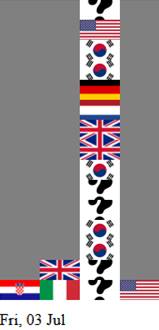}} 
\caption{\label{fig:gpbib_3-jul-2020}
Surge in 
\href{http://gpbib.cs.ucl.ac.uk/download/download.html}
{downloads} via the GP bibliography during the GI workshop
(14:00--17:21~BST).
(Each day divided into four 6~hour periods.)
}
\end{figure}

\section{Human Factors and GI}
There are now several implementations of genetic improvement in use
\cite{Langdon:2015:GECCO,langdon:2016:GI,Langdon:2016:GPEM,%
Langdon:2017:GI}
(e.g.\
Janus Manager~\cite{Haraldsson:2017:GI,Haraldsson:thesis}
and Facebook's
\mbox{SapFix}~%
\cite{Alshahwan:2019:GI,%
Jia:2018:SapFix,%
Marginean:2019:ICSE})
so it would be the right time to perform human studies to gain
insights from developers who have already used GI on what else would
be useful.
Notice all genetic improvement systems in use require human acceptance
of automatically generated source code changes.
Indeed in both 
Janus Manager (Python)
and \mbox{SapFix} (Hack, Java)
developer acceptance is explicitly part of the
continuous integration development protocol.

There is a great need to understand in more depth the human factors
surrounding the application and implementation of genetic improvement techniques within
industry.  
This should go beyond the consideration of the usability of
GI techniques to also examine how such techniques might affect
software developers’ workflow and work satisfaction. 
A broader focus would enable a richer understanding of the potential
barriers to GI techniques' adoption in industry, 
as well as the opportunities presented.

Concentrating on usability (often within the context of a controlled experiment) may also obscure the applicability of genetic improvement techniques in real-life settings, and what kind of specific techniques software developers would find useful in their work. However, usefulness alone is not enough and it must also be complemented with trustworthiness, since developers need to be able to trust GI techniques in order to use them with confidence. 

It is also vital to understand the broader impact of genetic improvement from a human perspective, such as considering what kind of tasks GI techniques might both remove from and add to software developers' workflow and what this might mean for the future of software engineering work.

\section{Explainabiloity,\\ Explain, Justify, Exploit}
\label{sec:explain} 

Explain, justify and exploit in GI go hand in hand. 
If how the algorithm works, 
and the resulting patches can be explained, 
their use can be justified to a larger community and to industry at large.
This in turn increases adoption of GI end opens it for exploitation.

\subsection{Explainable Genetic Improvement}

A very engaging discussion on explainable GI ensued during the workshop. 
Explainability is an important topic in the Artificial Intelligence~(AI) literature now
\cite{gunning2019darpa}. 
It provides users with the reasoning about decisions made during the
machine learning, search and optimisation process. 
It was suggested, that if we can provide explainability along with our patches, 
this would go a long way in helping users gain confidence in
the proposed program modifications. 
This fits in well with our earlier discussion of trust on the human
side of GI\@. 
Some argued that GI systems have {\em explainability built in}
as the algorithms explore parts of a search space and that it should
be possible to capture some of the reasoning 
(why some patches are selected or not) during the search. 
The discussion centered on two aspects of explainability:
structural and semantic.
Attendees felt that the semantic explainability is
harder and it would not be trivial to build in.  
Two potential issues are that 
(1) we are guiding the search towards the positive part of the search
space which could bias our explanations, 
and 
(2) since part of explainability is providing a minimally
explainable set this could lead to overfitting. 
Both of these issues would need to be overcome. 
All agreed that research into explainability of GI is an interesting
and exciting direction. %that some of us should pursue.

Through the discussion, we shared some research methodologies for
explainable GI\@.
One approach is to understand the GI process.
We already know the goals of the GI objective (fitness) function and which
code changes (mutations) %operators 
have been applied to the program. 
Based on the record of the changes, we hope to be able to explain how
the control flow of the program has changed 
and how performance (fitness) has changed after applying the change operators. 
While the approach of {\em understanding} the GI is a
passive strategy to achieve explainable GI, 
we can also actively {\em control} what GI can do. 
We may use modification %mutation 
operators that are designed to explain their changes. %semantically. 
Therefore, each change %mutation 
can provide more useful information 
about %for 
itself.

A more pragmatic approach could be to build on 
standard methods used for debugging, including
control flow and data flow graphs, which would aid developers,
if not to accept the patch then understand its 
impact on the code base. %influence. ???
Perhaps even if the fix is not the right one,
if developers understand its influence then 
they could still accept it as a temporary fix,
whilst they think of a permanent one. 
Also,
if we have a reasonable measure of readability,
GI could use it as one of the objectives 
in a multi-objective search.

\subsection{Justify}
The discussion turned towards the need for popular, easy-replicable use
cases of 
GI\footnote{%
Last year,
in Section~3 \cite{Langdon:2019:SEN},
we listed 22 tools,
also
\cite{Haraldsson:2020:GECCOcomp,%tutorial
Brownlee:2019:GECCO,Petke:2019:SSBSE,%GIN
an:2019:fse%PyGGI
}
were published after the GI @ ICSE 2019 workshop.
}.
It was suggested that in general people do not understand genetic algorithms 
and that perhaps this makes it a challenge to justify use
of GI to a broader audience.

During the discussion it was also noted that some form of metric
could be introduced for automatically generated patches.
Such a metric could also impact use of patches, similar to how upvotes
work in forums such as Stack Overflow.
In production environments,
such as continuous integration~(CI),
we certainly see that managers ensure that effectiveness metrics,
such as fraction of auto-patches accepted into production,
are collected
\cite{Alshahwan:2019:GI,%
Marginean:2019:ICSE}.

\subsection{Exploit}

The question of exploiting GI was felt to be 
much the same as
how to get uptake of GI
by industry
(see Sections~\ref{sec:mike_start} and~\ref{sec:wes_gi4}).

\section{Industry's View of GI}
\label{sec:mike_start} 
\subsection{Artificial Intelligence is very popular across Industry}

AI is one of the hottest topics in industry at the moment. 
Usually businesses that use the term AI refer to the technology that
will help them automate their current processes 
(e.g., chat bots,
automatic document text extraction, speech to text), 
better understand their data 
(machine learning clustering models) and 
take automatic prediction decisions 
(supervised learning models). 
Thus, there is a very big increase in hiring data scientists who
can apply the newest machine learning techniques. 
Knowledge of machine learning libraries such as Keras, Tensorflow and Pytorch% 
\footnote{See \url{https://keras.io/},
  \url{https://www.tensorflow.org/}, and \url{https://pytorch.org/}.}
has
become essential knowledge in most interviews. 
Practically, in industry the term AI refers to
machine learning related problems.

\subsection{Machine Learning has dominated the AI space}

The popularity of AI means that any problem that is in the
optimisation space is tackled usually as a machine learning problem. 
For instance, a lot of industrial problems that have to do
with parameter tuning for maximising or minimising an objective function,
are treated like machine learning prediction problems,
whilst they could be solved, many times, more efficiently using
Genetic Algorithms and other optimisation techniques. 
However, few people in industry are familiar with terms
such as genetic algorithms, 
multi-objective optimisation and genetic improvement. 
This may be due to limited industrial use so far
or not being highly advertised or
due to the lack of widely available and popular open source
libraries that are used by current data scientists.

\subsection{AI on data vs.\ GI on code adoption}

Most AI techniques in industry are applied on data, whereas %on the contrary,
most GI techniques focus on code
(recent GI exceptions include \cite{langdon:2018:EuroGP,Langdon:2018:SSBSE}). 
In many organisations, 
having access to data and exporting them for applying machine learning
libraries is quite a straightforward and replicable process. 
However, applying GI techniques on existing internal code bases can be
a bit more tricky because of the different programming languages used
(different teams use different languages inside the same company), 
the different programming tools, and the variety of project code
structures.

Additionally, many code bases lack proper testing and performance
benchmarks, upon which many GI techniques rely on. 
For instance, when we wanted to apply Artemis~%
\cite{Basios:2018:FSE}, %,DBLP:journals/corr/BasiosLWKLB17}, 
a GI tool that does automatic code optimisation using better data
structures on a performance critical component of a system, 
we realised that the project did not contain a proper performance
benchmark and relying on the test cases 
was not a realistic behaviour of the application, 
thus making the optimisations impractical. 
Additionally, building a performance benchmark for the specific
project would take a long time, and thus was considered a lower priority
on the 
project manager
list of features.

To summarise, the most common
issues when applying GI techniques on existing
industrial code bases are:
\begin{enumerate}
\item
Lack of proper testing and benchmarks on the code bases. Industrial
applications are usually much more complex and with many dependencies
than the applications used by the research community to apply GI.

\item
Lack of popular, easy-replicable successful use cases where GI has been
applied. Industry likes ready to use tools that can help them make
money. 

\item
GI tools are mostly understood by only a few people inside an
organisation, who are usually technical experts. Technical people have less
access to budget and investment in new technologies.

\item 
Further need to integrate GI as part of the continuous integration %~(CI)
process. If GI is inside a development tool, most engineers will not
focus on the exact techniques used, and will not be familiar with the
term GI.
\end{enumerate}

\label{sec:mike_last} 

\section{Researchers %Academe %Uni.\ working with 
and Industry}
\label{sec:wes_gi4}

In addition to TurinTech
(Sections~\ref{sec:mike_start}--\ref{sec:mike_last}),
there were 
two other companies represented in the registered audience: 
Facebook and GrammaTech. % and TurinTech.
The question of how GI researchers could work with them (or other companies)
was raised.
A Facebook call for research proposals was announced during the keynote.
For examples,
would industry accept partnership agreements?  
Perhaps these could lead to
easy ways to try out a new technique (e.g.,
stack-based GI, presented by Dr.~Blot~\cite{Blot:2020:GIsp}) 
on Facebook's SapFix\cite{Marginean:2019:ICSE}?

In order to develop GI techniques that are beneficial to industry,
greater collaboration with industry may be needed. 
However, there are several potential barriers to this. 
Firstly, some companies may be wary of sharing code and documentation
with researchers. 
(This can often be overcome with suitable 
non-disclosure agreements, NDAs.)
Secondly, the time investment needed for industry to work
with researchers might be off-putting. %in highly time-pressurised work
Several possible solutions were discussed,
including clearly articulated partnership agreements 
(traditional university ethics procedures may not be necessarily adequate~%
\cite{Rybnicek}) 
and adopting agile-inspired frameworks in order to carry out research in a
timely and light-touch fashion~%
\cite{Sharp14}.

Dr.\ Winter %One of the presentations Squeeze
\cite{Winter:2020:GI}
spoke of the need for more longitudinal and ethnographic research in industry,
so as to more fully understand the organisational and human context of
GI techniques' use, 
and potential barriers to their adoption.

More immediately,
a GI plug-in or simply %actually just 
better documentation will encourage %was to go to to my date 
way more widespread use of GI.

\begin{table*}[tb]
\caption{
\label{tab:erik}
Contrasting Approaches and Tool Usage between GI and SBST
ICSE~2020 workshops
}
\begin{center}
\begin{tabular}{|l|l|l|}
\cline{2-3}
\multicolumn{1}{c|}{} %Tool 
&
Genetic Improvement &
\multicolumn{1}{c|}{Search-Based Software Testing}
\\
\hline
Meeting Software &
Zoom &
Microsoft Teams
\\
Livestream &
YouTube &
Twitch
\\
Presentation
Format &
Live presentation &
Live keynote + pre-recorded \mbox{author} videos
\\\hline
\end{tabular}
\end{center}
\end{table*}
 %nope have to move table by hand \clearpage

\vspace*{-2ex} %Squeeze
\section{Benchmarks}

A call for industry cooperation to create realistic genetic improvement benchmarks was
raised. Some argued that benchmarks which have characteristics of real
(industrial) applications are needed to allow researchers to build new
techniques that will work in practice.  
If researchers use only small benchmarks, which are open source, their
techniques may overfit and/or not scale to real systems.  
Another benefit of benchmarks is to provide an easy entry point for
newcomers to GI\@. 
Some of the attendees from industry at the workshop pushed back saying
that everyone already has access to realistic systems since many of the
large open source applications today are representative of the types
of systems in industry.  
However, they also noted that to make research techniques viable for
industry they need to be built into existing 
continuous integration (CI) build systems. 
They suggested GitHub open source repository pull requests 
be provided 
for genetic improvement patches to find out if
they are really good and useful. %patches. 
Some suggested that scalability, especially in data-driven systems, is
still a real challenge for genetic improvement. 
Others argued that some GI tools scale well, but that this may be a
factor of whether or not there is good fault~(bug) localization behind it.

Benchmarking is a challenge that even the
optimisation community continues to struggle with
\cite{bartzbeielstein2020benchmarking}. %Markus
Indeed
the fact that the ten-year impact award at GECCO 2020 was awarded to a
paper on benchmarking organisation \cite{Hansen:2010:geccocomp},
highlights that the challenge is still unsolved. %and is common across
If real world systems taken from industry might provide
realistic examples, they come with their inherent cost and
complexity.
Furthermore, when selecting representatives examples: which
project to consider, which programming language to target, what
feature in general to take into consideration, are recurrent
questions in the genetic improvement community for which there is no clear
consensus.
One idea,
in order to obtain much simpler and cheaper problems
instances for GI approaches,
would be to take a step back and from the theory point of
view focus on the core components and characteristics of GI problems~% 
\cite{Blot:2020:GI}.
Such theory-anchored characterisations, while not exactly filling
the need of actual complex real-world benchmarks, would have
several advantages: 
\begin{enumerate}
\item
better rationalisation of what
makes GI work, 
\item
minimal examples highlighting specific features for
practitioners to improve the state-of-the-art on, 
and 
\item
an opportunity 
for people from more traditional optimisation fields
to encounter GI problems and share their expertise.
\end{enumerate}
Finally,
to further
strengthen %synergise 
the connection 
between the software engineering roots of GI and the
knowledge of the optimisation community, 
GI might provide benchmarks for
optimisation competitions 
(e.g., held at the CEC or GECCO conferences).

\section{Next Year:\\ What if we still have a Pandemic}
\label{sec:2021}

\subsection{Workshop Virtualization}

A new issue that all international gatherings had to manage this year was
remote interaction and content delivery for participants located in
varying 
\href{http://www.cs.ucl.ac.uk/staff/W.Langdon/egp2020}
{time zones}.  
While ICSE2020 used a time band model, both 
GI~@~ICSE~2020
and 
the Search-Based Software Testing (SBST~2020)
workshops 
had single synchronous sessions using an online meeting tool
and accompanying livestream.  Table~\ref{tab:erik} summarises the
two approaches. %differences between the two events:

In terms of ``success'', both workshops were enjoyable to attend and
available technologies minimized problems during both events.  The key
difference between the two formats was that GI was more
presentation-focused and SBST was more discussion-focused.  For SBST,
additional effort was placed on the chairs to ensure that there is no
``dead air'' (i.e., questions needed to be prepared in advance). 
Whereas
GI was able to leverage the authors as well as the audience
to fuel the interaction.
SBST also provided a social event in the form of an enjoyable 
live pub quiz.
Unlike other international events, %GECCO + CEC
the ICSE workshops were {\em not} under heavy preasure to
remain within their scheduled time slot
or under the threat of Zoom being closed
whilst discussion continued.

Regardless of the differences, both GI and SBST were both successful
in their own right.  %An interesting point to note is that 
Both had a
record number of registered participants. 
Monitoring attendance rates in the future will determine
whether this is a result of lower attendance fees, being
available via the world wide web (WWW), or both.
However this initial metric makes a strong case
for allowing virtual participation via the Internet in future. %iterations.

\subsection{Do What the Virtual World Does Best}

Perhaps we should not try to
replicate the conference experience 
but work out what is best done in the virtual world 
and more importantly how
to connect via it to give faster discussion
(rather than the current yearly or bi-annual cycle).

The use of 
\href{https://slack.com/}
{Slack} to continue conversations after scheduled
talks/sessions had finished was something that worked well at other
conferences %(e.g.\ ICT4S) 
and enabled more informal interaction.

\bibliographystyle{unsrturl}

\balance
{\footnotesize %approx 9pt

\bibliography{gi2020_discussion}
}

\end{document}